\documentclass[%
 10pt,
 amsmath,
 amssymb,
 aps, 
 pra,
 twocolumn,
 longbibliography
]{revtex4-2}

\usepackage[utf8]{inputenc}
\usepackage{fontenc}
\usepackage{graphicx}
\usepackage{bm}
\usepackage{hyperref}
\usepackage{tikz} 
\usepackage[caption=false]{subfig}
\usetikzlibrary{quantikz2}

\begin{document}


\title{Variational-Adiabatic Quantum Solver for Systems of Linear Equations with Warm Starts}

\author{Claudio Sanavio}
\affiliation{Italian Institute of Technology, Viale Regina Elena 291, 10161, Rome, Italy}
\email{claudio.sanavio@iit.it}
 
\author{Fabio Mascherpa}
\affiliation{HPCOX/A – Advanced Modeling and Simulation, Eni S.p.A., DIT – Digital  \& Information Technology, Via Emilia 1, 20097 San Donato Milanese (MI), Italy}
\email{fabio.mascherpa@eni.com}

\author{Alessia Marruzzo}%
\affiliation{HPCOX/A – Advanced Modeling and Simulation, Eni S.p.A., DIT – Digital  \& Information Technology, Via Emilia 1, 20097 San Donato Milanese (MI), Italy}

\author{Alfonso Amendola}%
\affiliation{HPCOX/A – Advanced Modeling and Simulation, Eni S.p.A., DIT – Digital  \& Information Technology, Via Emilia 1, 20097 San Donato Milanese (MI), Italy}

\author{Sauro Succi}%
\affiliation{Italian Institute of Technology, Viale Regina Elena 291, 10161, Rome, Italy}%

\date{\today}

\begin{abstract}
We propose a revisited variational quantum solver for linear systems, designed to circumvent the barren plateau phenomenon by combining two key techniques: adiabatic evolution and warm starts. To this end, we define an initial Hamiltonian with a known ground state which is easily implemented on the quantum circuit, and then “adiabatically” evolve the Hamiltonian by tuning a control variable in such a way that the final ground state matches the solution to the given linear system. This evolution is carried out in incremental steps, and the ground state at each step is found by minimizing the energy using the parameter values corresponding to the previous minimum as a warm start to guide the search. As a first test case, the method is applied to several linear systems obtained by discretizing a one-dimensional heat flow equation with different physical assumptions and grid choices. Our method successfully and reliably improves upon the solution to the same problem as obtained by a conventional quantum solver, reaching very close to the global minimum also in the case of very shallow circuit implementations.
\end{abstract}

\maketitle


\section{\label{sec:I} Introduction}

Linear systems of equations can be used to describe countless physical phenomena, ranging from transport models to linear fluid dynamics or optimization processes in logistics~\cite{moin_fundamentals_2010}. Linear algebraic systems can arise from the discretization of linear partial differential equations (PDEs) in space and time, and even nonlinear PDEs can be reduced to a linear system by employing suitable linearization techniques, such as Taylor expansion or Carleman linearization as applied to the lattice Boltzmann equation~\cite{LatticeBoltzmann1992,itani_analysis_2022,sanavio_three_2024,sanavio_explicit_2025}. Given the ubiquity of such mathematical problems and the possibly very large complexity of their numerical solution, the relevance of devising efficient methods for this task can hardly be overstated. A straightforward method for solving a linear system of equations involves brute-force inversion, for instance via Gaussian elimination, but this method suffers from numerical instability and a computational complexity of order $\mathcal{O}(N^3)$ in the dimension $N$ of the problem. Faster methods, such as Krylov subspace--based or iterative relaxation methods, can reduce this cost significantly (even down to order $\mathcal{O}(N)$) but introduce some degree of approximation and controlled error~\cite{kaneda_brief_2007}.

In the last decade, quantum computing for the solution of linear systems has become an emerging field of study, driven by recent technological advances and the promise of a possibly exponential speedup in several different fields, from the simulation of quantum systems to the solution of mathematical models with applications e.g.\ in engineering, fluid dynamics and geophysics~\cite{succi_quantum_2023,sanavio_quantum_2024,basin_modeling}. Among these, in 2008 the Harrow-Hassidim-Lloyd (HHL) algorithm~\cite{HHL} showed how linear systems could be solved by a quantum computer up to an error $\epsilon$ with a computational complexity of order $(\log N)/\epsilon$, later improved to $\log(N/\epsilon)$ by Childs et al.~\cite{berry_exponential_2014, childs_quantum_2017}. The latter method, further improved with quantum signal processing techniques~\cite{low_optimal_2017}, currently represents the state-of-the-art quantum algorithm for solving a linear system of equations. Unfortunately, while theoretically excellent, this algorithm involves large numbers of logical quantum gates and therefore requires a fault-tolerant quantum computer to give meaningful results. Despite the significant improvements achieved by recent developments in quantum hardware design and operation, current machines do not yet support the levels of error correction needed to suppress all the quantum noise involved in any reasonably sized implementation of the method~\cite{campbell_series_2024}.

For this reason, most research efforts in recent years have shifted to algorithms that could turn out to be useful on so-called Noisy Intermediate-Scale Quantum (NISQ) devices, which comprise hundreds of qubits and noisy quantum gates. Among such methods, hybrid variational algorithms~\cite{cerezo_variational_2021} have been attracting particular attention. In a variational quantum linear solver (VQLS)~\cite{bravo-prieto_variational_2023,gnanasekaran_efficient_2024}, one constructs a quantum circuit whose Hilbert space is large enough to accommodate the original problem (i.e.\ the number of qubits $n$ needs to be such that $2^n\geq N$) and acts on those qubits with a given operator $U(\theta)$, known as an \textit{ansatz}, typically consisting of one- and two-qubit gates and controlled by a set of parameters $\theta$: one then defines a cost function such that its minimum with respect to $\theta$ corresponds to a state $\ket{x(\theta)}$ representing the solution of the linear system.

However, even variational quantum algorithms, although promising due to their logarithmic use of resources and potential applications that venture into the field of quantum machine learning~\cite{biamonte_quantum_2017,schuld_machine_2021}, suffer from a serious drawback in the form of the so-called Barren Plateaus (BP)~\cite{cerezo_variational_2021,bravo-prieto_variational_2023,cerezo_cost_2021,cerezo_challenges_2022,larocca_diagnosing_2022,BP_Nature,BP_Review2025}. In fact, as the number of qubits increases, the gradient of the cost function becomes smaller in magnitude throughout most of the parameter space. Crucially, this decrease turns out to be exponential in the number of qubits, i.e.\ for any small number $\epsilon>0$ the probability of finding a set of parameters $\theta$ such that $|\nabla_\theta f|>\epsilon$ decreases as $P_\theta(|\nabla_\theta f|\geq\epsilon)\sim\mathcal{O}(2^{-n})$. In such regions, both gradient-based and global optimizers are unable to minimize the cost function efficiently, defeating the purpose of exponentially reducing the problem size in the first place.

The BP problem has been studied extensively, with various approaches and ideas on how to avoid or mitigate it~\cite{BP_Nature,BP_Review2025}, but it appears to be a built-in feature of the variational quantum algorithm family. In particular, it is worth mentioning one recent study~\cite{cerezo_2025}, which uncovered a major pitfall of using local cost functions, i.e. cost functions computed from properties of individual qubits or small groups of qubits which do not grow with the system size. Local cost functions were considered appealing for variational methods, because they are better behaved than global ones and may not exhibit BPs~\cite{cerezo_cost_2021}. However, in Ref.~\cite{cerezo_2025}, the authors highlighted a link between the improved scaling properties of variational algorithms employing such cost functions and the reduction in complexity of the variational problem itself if formulated this way, and showed that any problem that can be solved by a variational method with local cost function is also efficiently solvable by a classical algorithm. 
To put it bluntly, formulating a variational algorithm with a local cost function removes both the BP and any quantum advantage.

In this work, we attempt a different route around this hurdle, building upon two ideas already tested in the context of quantum Hamiltonian simulation. First, following Ref.~\cite{subasi_quantum_2019}, we redefine the variational problem as a problem family parametrized by a guiding variable, which is used as a control to gradually move from a trivial linear system to the one of interest in a fashion similar to the principle of adiabatic quantum computing~\cite{costa_optimal_2022,jennings_randomized_2025}. The other key ingredient of the proposed approach is the well-known concept of warm starts in variational methods, described e.g. in Refs.~\cite{egger_warm-starting_2021,guerrero_bee-yond_2025,puig_variational_2025}: we use available information on the problem to initialize the search for a minimum in a region with a sufficiently steep descent path towards it---in other words, outside the BP---by changing the parameter landscape in steps small enough to follow the minimum as it shifts from the initial, trivial solution to the one of the problem at hand.

In Section~\ref{sec:II}, we will describe the proposed approach in detail. In Section~\ref{sec:III}, we will demonstrate an example application by solving linear systems derived from a finite-difference formulation of the heat flow equation in one spatial dimension using different physical settings of the problem. Finally, in Section~\ref{sec:IV}, we will discuss the quality of the results and the performance of the algorithm, and present our conclusions and outlook.

\section{\label{sec:II} Adiabatic variational solver with warm starts}

Given an invertible linear system 
\begin{equation}\label{eq:classical_linear_system}
    A\mathbf{x}=\mathbf{b},
\end{equation}
we can employ the standard VQLS~\cite{bravo-prieto_variational_2023} to find the solution of the corresponding quantum linear system
\begin{equation}\label{eq:quantum_linear_system}
    A\ket{x}\propto\ket{b},
\end{equation}
or in inverted form,
\begin{equation}\label{eq:QLS_solution}
    \ket{x}=\frac{A^{-1}\ket{b}}{||A^{-1}\ket{b}||},
\end{equation}
by expressing Eq.~\eqref{eq:classical_linear_system} in the language of quantum states and operators defined on a suitable parametric quantum circuit, and then solving for the parameters. From this point on, we will assume the system Eq.~\eqref{eq:classical_linear_system} to be $N=2^n$-dimensional, padding $\mathbf{x}$ and $\mathbf{b}$ with zeros and $A$ with the identity matrix on any dimensions not present in the original system given. The equivalent quantum system~\eqref{eq:quantum_linear_system} is defined by normalizing the vectors $\mathbf{x}$ and $\mathbf{b}$ to unity, so that they can be interpreted as quantum states $\ket{x}$ and $\ket{b}$, and relaxing the equality since $A$ is not necessarily unitary.

In a variational setting, the state $\ket{x(\theta)}$ is obtained by applying a parametric ansatz $U(\mathbf{\theta})$ to the initial state of the quantum circuit:
\begin{equation}
    \ket{x(\mathbf{\theta})} = U(\mathbf{\theta})\ket{0},
\end{equation}
and the VQLS identifies the correct values of the parameters to solve Eq.~\eqref{eq:quantum_linear_system} by minimizing the appropriate cost function, according to the variational principle
\begin{equation}\label{eq:VarPrinciple}
    C(\mathbf{\theta}^*)=\min_{\mathbf{\theta}} C(\mathbf{\theta})\iff \ket{x(\theta^*)}\approx\ket{x}.
\end{equation}

Note that the approximate relation above is a consequence of the so-called \textit{expressibility} of the ansatz $U(\theta)$~\cite{Expressibility}: in general, $\ket{x(\theta^*)}$ will be the closest possible state to the real solution $\ket{x}$ among those accessible by $U$, which does not necessarily contain $\ket{x}$ itself.

Although elegant, the VQLS as it is offers little practical value, since optimizing a global cost function is generally a hard task due to the cost function landscape taking the form of a BP. Nevertheless, previous works have highlighted both its wide range of potential applications and the quality of the results obtained either for small-scale problems or for more complex ones which allow for local cost functions and efficient classical solutions.

\subsection{\label{ssec:AVQLS}Adiabatic variational solver}

In order to address the challenges presented by the standard VQLS and to devise a strategy capable of working around them, we considered a variation of the adiabatic quantum solver introduced in Ref.~\cite{subasi_quantum_2019}, which defines a parametric version of the linear system Eq.~\eqref{eq:quantum_linear_system} and uses a stochastic evolution simulation to solve it adiabatically. Rather than following that path, we will borrow the idea of an adiabatic parametrization of the problem and integrate it into the VQLS.

First, let $A$ be a non-singular $N\times N$ matrix with $N=2^n$ and define the parametric linear system
\begin{equation}\label{eq:QLP_s}
    A(s)\ket{x(s)}\propto\ket{\bar{b}},
\end{equation}
where
\begin{eqnarray}
\label{eq:A_pos}
    A(s)&\coloneqq&(1-s)\mathbb{I}+sA,\nonumber\\
    \ket{\bar{b}}&\coloneqq&\ket{b}
\end{eqnarray}
if $A$ is positive, and
\begin{eqnarray}
\label{eq:A_not_pos}
    A(s)&\coloneqq& (1-s)\sigma_z\otimes\mathbb{I}+s\sigma_x\otimes A,\nonumber\\  
    \ket{\bar{b}}&\coloneqq&\ket{+}\otimes\ket{b}
\end{eqnarray}
if it is not positive. The latter definition ensures that $A(s)$ is invertible for any $s$ (and hence that the corresponding solution $\ket{x(s)}$ exists), and implies adding an ancilla qubit to perform the algorithm.

The key of the parametric description of the problem is that at $s=0$ one immediately finds that $\ket{x(0)}=\ket{b}$ or $\ket{x(0)}=\ket{-}\otimes\ket{b}$, depending on the positivity of $A$, while at $s=1$ the solution is $\ket{x(1)}=\ket{x}$ or $\ket{x(1)}=\ket{+}\otimes\ket{x}$. Therefore, the parametrization in $s$ allows us to interpolate smoothly between the trivial linear problem with $A(s)$ proportional to the identity and the actual problem defined by Eq.~\eqref{eq:quantum_linear_system}.

Furthermore, it can be readily shown that for each $s$ the solution $\ket{x(s)}$ of Eq.~\eqref{eq:QLP_s} is also the ground state of the Hamiltonian
\begin{equation}\label{eq:H_s}
    H(s)\coloneqq A^\dagger(s)\left(\mathbb{I}-\Pi_b\right)A(s),
\end{equation}
where $\Pi_b=\ket{\bar{b}}\bra{\bar{b}}$ is the projector on the state $\ket{\bar{b}}$, and that this ground state is unique and has eigenvalue $0$. Therefore, combining all these elements together, one may treat the problem as an adiabatic search for the ground state of the Hamiltonian $H(s)$ as $s$ gradually increases from $0$ to $1$~\cite{subasi_quantum_2019}.

The problem can then be approached variationally by simply defining a cost function family parametrized by $s$, which corresponds to the energy expectation value
\begin{equation}\label{eq:CostF}
    C_s(\theta)\coloneqq\bra{0}U^{\dagger}(\theta)H(s)U(\theta)\ket{0}.
\end{equation}
The solution $\ket{x}$ is obtained by first solving the trivial problem at $s=0$ and then letting the parameters follow the minimum as $s$ changes. This raises two questions: how do we find the first minimum, and how do we discretize the smooth parametrization into steps small enough to ``track'' the minimum as it moves through the parameter space. 

The first question can be ignored if one assumes an oracle capable of preparing arbitrary quantum states; otherwise, one might suspect that finding the correct parameters to construct the state $\ket{b}$ may be just as hard as solving the original problem. Fortunately, this is not the case: it is always possible to apply a unitary transformation $S$ to the linear system Eq.~\eqref{eq:classical_linear_system} such that $\mathbf{b}$ is mapped onto a vector with a simple expression in terms of the ansatz parameters, such as the first standard basis vector $\mathbf{e}_1=(1,0,0,\dots)^t$:
\begin{equation}\label{eq:householder_system_equation}
     S AS^\dagger S\mathbf{x}=\tilde A \tilde{\mathbf{x}} = \mathbf{e}_1,
\end{equation}
where $\tilde A$ and $\tilde{\mathbf{x}}$ are the transformed matrix and solution. One convenient choice for $S$ is known as the Householder matrix~\cite{householder_unitary_1958}, which is defined as follows: first subtract $\mathbf{e}_1$ from the normalized $\mathbf{b}$ to define a vector $\mathbf{v}\coloneqq\mathbf{b}/||\mathbf{b}||-\mathbf{e}_1$, then use this to define a reflection matrix as
\begin{equation}\label{eq:householder_matrix}
    S_\mathrm{H}\coloneqq\mathbb{I}-2\frac{\mathbf{v}\mathbf{v}^\dagger}{\mathbf{v}\cdot\mathbf{v}}.
\end{equation}
It can be easily checked that $S_\mathrm{H}$ is unitary, Hermitian and involutory, i.e.\ $S=S^\dagger=S^{-1}=S$. From now on, and without loss of generality, we will assume the linear system to be already expressed in this basis, i.e.\ $\mathbf{b}=\mathbf{e}_1$, or $\ket{b}=\ket{0}$ on the quantum circuit.

It should be noted that although this operation is not computationally cost-free, it shifts some of the burden from the quantum circuit to the classical pre-processing phase. We remark that the quantum device is not assumed to provide any computational advantage in the preparation of classical states but only in the resolution of well-defined linear problems~\cite{bravo-prieto_variational_2023,VQLS_Turati}, so we chose to phrase the problems in a suitable form before feeding them to the quantum algorithm. As the definition Eq.~\eqref{eq:householder_matrix} suggests, the Householder matrix itself is easily obtained from the classical vector $\mathbf{b}$ at a cost directly linked to that of acquiring it from memory.

The second question is far from trivial and will be the focus of the next subsection, where we will introduce the concept of warm starts.

\subsection{\label{ssec:WS}Improving efficiency by using warm starts}

\begin{figure}
    \centering
    \includegraphics[width=1\linewidth]{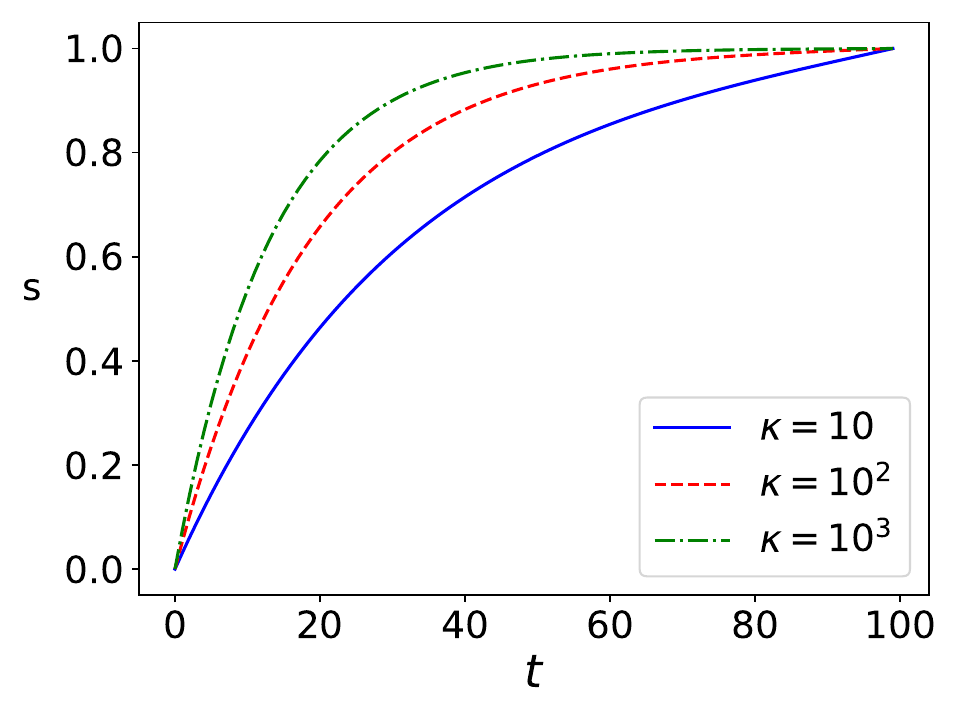}
    \caption{Increase of $s$ as a function of $T=100$ values of $v$ for condition numbers $\kappa = 10$ (solid blue line), $\kappa=10^2$ (dashed red line) and $\kappa = 10^3$ (dot-dashed green line). Note the concentration of $s(v)$ at values closer to $1$ for higher $\kappa$.}
    \label{fig:schedule_somma}
\end{figure}

\begin{figure*}[htbp]  
    \centering
    \includegraphics[width=0.8\linewidth]{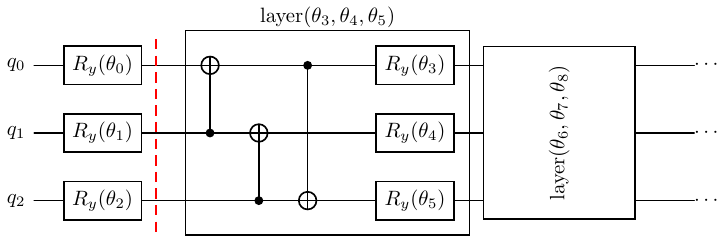}
    \caption{Structure of the parametric quantum circuit $U(\theta)$ for the case $n=3$. The circuit starts with a $R_y$ rotation for each qubit, followed by $d=2$ repetitions of the layer detailed in the first box, which combines more single-qubit rotations and a set of two-qubit $\text{CNOT}$ gates arranged in a ring configuration.}
    \label{fig:parametric_ansatz}
\end{figure*}

The question of how to increase $s$ by a controlled amount $\delta s$ such that the minimum of the new cost function $C_{s+\delta s}(\theta)$ can be found efficiently knowing the previous minimum $C_s(\theta^*)$ is a subtle one. We address this point by resorting to the concept of warm starts, i.e.\ the idea that one can avoid barren plateaus by initializing the search for a minimum at a point close enough to the global minimum for the gradient of the cost function to be non-negligible.

Recent works~\cite{puig_variational_2025,mhiri_unifying_2025} have proved some rigorous links between a change in the cost function and the distance between the minima before and after the change, in the context of Hamiltonian evolution simulation. In particular, in Ref.~\cite{puig_variational_2025} it is shown that the evolution of a quantum state may be simulated variationally with rigorously bounded warm starts by letting the system evolve in timesteps short enough for the cost function (which in that problem is an infidelity measure $\mathcal{I}_{\delta t}(\theta)= 1 -|\bra{\psi(\theta)}e^{-iH\delta t}\ket{\psi(\theta^*)}|^2$) at the previous minimum $\theta^*$ to descend unambiguously towards the new one at each step. In the following paragraphs, we will show how we applied a similar concept to our adiabatic solver, for which no such rigorous bounds were proven, in a more heuristic way.

\begin{figure}
    \centering 
    \includegraphics[width=1\linewidth]{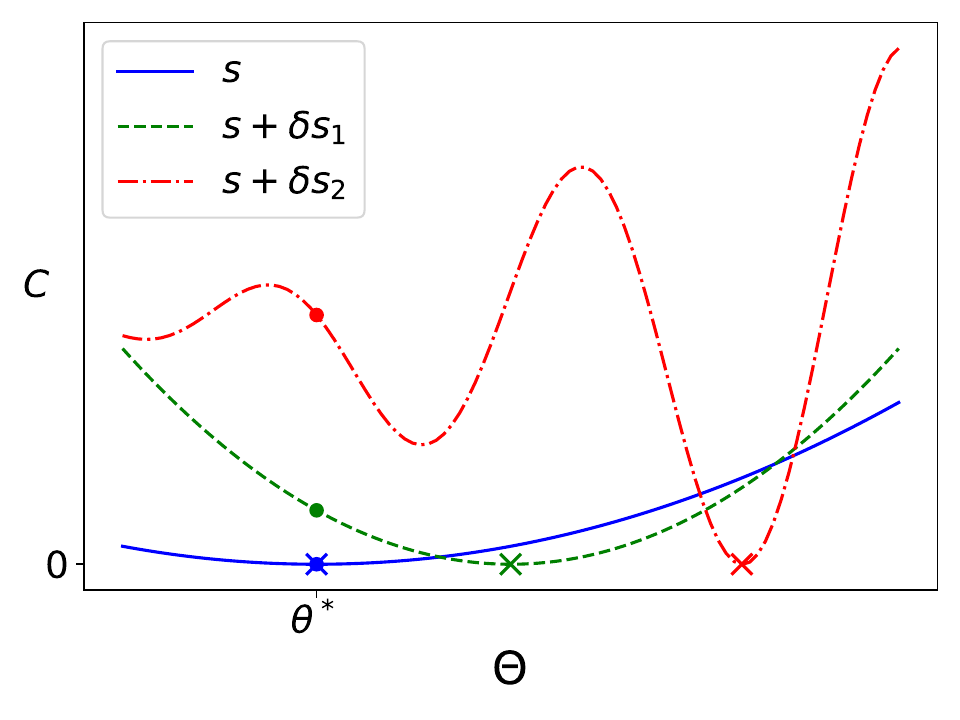}
    \caption{A slice of the energy landscape along a preferred direction $\Theta$: at a given $s$, the blue cross locates the minimum of the cost function (solid blue line). If we evolve the system by $\delta s_1$, the new global minimum (green cross) is still in the same valley as the previous one (dashed green line), but if we evolve it by $\delta s_2$ we overshoot the convexity region and the new global minimum (red cross) appears in another valley (dot-dashed red line), with no descending path joining it to the previous minimum (red dot).}
    \label{fig:energy_landscape} 
\end{figure}
First, any procedure to determine the step size $\delta s$ should take into account the fact that, depending on the condition number $\kappa$ of the matrix $A$, the rate of change of $H(s)$ and $\ket{x(s)}$ can be expected to increase as $s$ approaches $1$. Therefore, splitting the interval between $0$ and $1$ into a number of identical steps is not an efficient strategy to track the adiabatic transformation of the problem. A more suitable prescription which accounts for $\kappa$ was introduced in Ref.~\cite{subasi_quantum_2019}: let $v\in[v_\mathrm{min},v_\mathrm{max}]$, where
\begin{eqnarray}
        v_\mathrm{min}&=&\sqrt{\frac{2\kappa^2}{1+\kappa^2}}\log\left(\kappa\sqrt{1+\kappa^2}-\kappa^2\right), \\
        v_\mathrm{max}&=&\sqrt{\frac{2\kappa^2}{1+\kappa^2}}\log\left(\sqrt{1+\kappa^2}+1\right), \nonumber
\end{eqnarray}
and define the following dependence of $s$ on $v$:
\begin{equation}\label{eq:s_from_v}
    s(v)\coloneqq\frac{e^{v\sqrt{\frac{1+\kappa^2}{2\kappa^2}}}+2\kappa^2-\kappa^2e^{-v\sqrt{\frac{1+\kappa^2}{2\kappa^2}}}}{2(1+\kappa^2)}.
\end{equation}
It can be easily checked that $s(v)$ increases monotonically and that $s(v_\mathrm{min})=0$ and $s(v_\mathrm{max})=1$. Applying Eq.~\eqref{eq:s_from_v} to a set of evenly spaced values $v_j=v_\mathrm{min}+\frac{j}{T}(v_\mathrm{max}-v_\mathrm{min})$, with $j$ running from $0$ to some maximum integer $T$, one obtains an adiabatic parametrization that tilts the distribution of $s_j$ towards $1$ by an amount depending on the problem at hand. The specific dependence expressed by Eq.~\eqref{eq:s_from_v} is obtained by requiring that $||\partial_v\ket{x(s(v))}||\leq 1$ for all $v$ (the details of the derivation may be found in Ref.~\cite{subasi_quantum_2019}). In Fig.~\ref{fig:schedule_somma}, we show the nonlinear progression of $s$ using $T=100$ points for different values of $\kappa$: note how the points tend to cluster more and more near the upper limit as $\kappa$ increases.

The other element determining the cost function landscape and its dependence on $s$ is, of course, the structure of the ansatz itself. We chose to adopt the commonly used alternating ansatz introduced in Ref.~\cite{bravo-prieto_variational_2023}, which consists of a number $d$ of layers each comprising a rotation of each qubit by an independent angle and a loop of two-qubit entangling gates (either $\text{CNOT}$ or $\text{CZ}$), as shown in Fig.~\ref{fig:parametric_ansatz}. From now on, for the sake of simplicity, we will assume $A$ to be positive, so the circuit size will always be $n=\log_2N$; the whole treatment is easily generalized beyond positive matrices by introducing an ancilla qubit and following the definitions given in Eq.~\eqref{eq:A_not_pos}.

The cost function $C_s(\theta)$ for such a circuit has some useful properties, which will allow us to improve the adiabatic progression defined above and take full advantage of the warm-start approach. To see how, let us introduce some new notation in order to make the dependence of $C_s(\theta)$ on $s$ explicit: define
\begin{eqnarray}\label{eq:abc}
    \hat{a} &\coloneqq& \Pi_b-A\Pi_b-\Pi_bA+A\Pi_bA,\nonumber\\
    \hat{b} &\coloneqq& -2\Pi_b+A\Pi_b+\Pi_bA,\\
    \hat{c} &\coloneqq& \Pi_b,\nonumber
\end{eqnarray}
so that Eq.~\eqref{eq:H_s} may be rewritten as
\begin{equation}
    H(s) = s^2\hat{a}+s\hat{b}+\hat{c}
\end{equation}
by grouping the terms by powers of $s$. Plugging this expression for $H(s)$ into the definition of the cost function Eq.~\eqref{eq:CostF}, $C_s(\theta)$ can be expanded in powers of $s$ as well, with each coefficient equal to the appropriate expectation value on the quantum circuit state (denoted here by the shorthand $\langle O\rangle_\theta=\bra{0}U^{\dagger}(\theta)OU(\theta)\ket{0}$ for conciseness):
\begin{equation}\label{eq:C_expanded_s}
    C_s(\theta) = \langle H(s)\rangle_\theta = s^2\langle\hat{a}\rangle_\theta+s\langle\hat{b}\rangle_\theta+\langle\hat{c}\rangle_\theta.
\end{equation}
This allows one to calculate the value of the cost function at this particular point in parameter space for any \textit{other} value of $s$ as well: adding an arbitrary increment $\delta s$ in Eq.~\eqref{eq:C_expanded_s} and grouping the terms by powers of $\delta s$, one immediately finds
\begin{equation}\label{eq:cost_later}
    C_{s+\delta s}(\theta) = \delta s^2\langle\hat{a}\rangle_\theta+\delta s(2s\langle \hat{a}\rangle_\theta+\langle \hat{b}\rangle_\theta)+C_s(\theta).
\end{equation}
Note that this expression is exact for any $\delta s$.

The dependence of $C_s(\theta)$ on the ansatz parameters, on the other hand, is such that its derivatives can be computed exactly by employing the parameter-shift rule~\cite{ParamShift}: in particular, the gradient $\nabla[C_s(\theta)]$ and Hessian $\mathcal{H}[C_s(\theta)]$ are given by
\begin{eqnarray}\label{eq:PS_rule}
    \nabla_i C_s(\theta) &=& \frac{1}{2\sin(\beta)}\left(C_s(\theta+\beta\mathbf{e}_i)-C_s(\theta-\beta\mathbf{e}_i)\right),\\
    \mathcal{H}_{ij}C_s(\theta) &=& \frac{1}{2\sin(\beta)}\left(\nabla_jC_s(\theta+\beta\mathbf{e}_i)-\nabla_jC_s(\theta-\beta\mathbf{e}_i)\right),\nonumber
\end{eqnarray}
for any $\beta$, so that two evaluations of the cost function in $\theta\pm\beta\mathbf{e}_i$ give an exact gradient component and four evaluations in $\theta\pm\beta\mathbf{e}_i\pm\beta\mathbf{e}_j$ (or two in $\theta\pm2\beta\mathbf{e}_i$ plus $C_s(\theta)$ itself for $i=j$) give an exact Hessian matrix element.

Combining this with Eq.~\eqref{eq:cost_later}, we can use $C_s(\theta)$ to construct the Hessian $\mathcal{H}[C_{s+\delta s}(\theta)]$ for any given $\delta s$:
\begin{eqnarray}\label{eq:Hessian_expand_s}
    \mathcal{H}[C_{s+\delta s}(\theta)] &=& \delta s^2\,\mathcal{H}[\langle\hat{a}\rangle_\theta]
    +\delta s\,\big(2s\mathcal{H}[\langle\hat{a}\rangle_\theta]+\mathcal{H}[\langle\hat{b}\rangle_\theta]\big)\nonumber \\
    &&+\mathcal{H}[C_s(\theta)],
\end{eqnarray}
where each matrix element can be computed by applying parameter shifts as described in Eq.~\eqref{eq:PS_rule} above.

This may be used as a guide to determine the optimal $\delta s$ for a warm start. To see how, suppose that we found the global minimum of the cost function for $s=s^*$ at $\theta=\theta^*$, i.e.\ $C_{s^*}(\theta^*)=0$. By definition, $\mathcal{H}[C_{s^*}(\theta^*)]$ has only nonnegative eigenvalues at this point. In order to increment $s$ as much as possible without the new cost function $C_{s^*+\delta s}(\theta)$ becoming too flat at $\theta=\theta^*$, one can use Eq.~\eqref{eq:Hessian_expand_s} to find the largest possible $\delta s$ such that no eigenvalue of $\mathcal{H}[C_{s+\delta s}(\theta^*)]$ becomes negative, i.e.\ such that $\theta^*$ is still within the same trough or valley as the new minimum to be found. Fig.~\ref{fig:energy_landscape} shows a one-dimensional example of how a function may change and its minimum may shift, either remaining within the same convex region as the previous minimum or moving out of reach from it.

Note that this solution of a nonlinear system does not impact the overall complexity of the algorithm, because $\mathcal{H}[C_{s}(\theta)]$ is an $n_p\times n_p$ symmetric matrix, where $n_p=n(d+1)$ is the number of ansatz parameters and depends on the number of qubits $n$ and the circuit depth (i.e.\ the number of layers in the ansatz) $d$. Computing the eigenvalues of $\mathcal{H}[C_{s+\delta s}(\theta)]$ therefore has a cost of order $\mathcal{O}(n_p^3)$. The optimal $\delta s$ can be computed using iterative Newton-like methods, which determine $\delta s$ up to some error $\epsilon$ in $\mathcal{O}(\log(\log(1/\epsilon))$ steps: the total complexity of finding $\delta s$ up to this error is therefore of order $\mathcal{O}(n_p^3\log(\log(1/\epsilon)))$.

Of course, there is no guarantee that a numerical optimization will actually find a global minimum of the cost function. It could find a local minimum, or the best possible solution compatible with the ansatz, or it could return a saddle point. Therefore, it is necessary to check the Hessian at the current $s$ before moving on to the calculation of the next one through the procedure described above. We combined this strategy with the predetermined sequence of $s$ values given by Eq.~\eqref{eq:s_from_v}, using the latter as a fallback option whenever the $\delta s$ suggested by the Hessian calculation was not suitable for a warm start or the starting point was not a minimum.

More specifically, depending on the result of this convexity analysis, there are four possibilities at each step in the adiabatic variable $s$.
\begin{itemize}
    \item $\mathcal{H}[C_{s^*}(\theta^*)]$ not positive: we are not in a minimum. Increment to next $s$ suggested by Eq.~\eqref{eq:s_from_v}.
    \item $\delta s<0$: all positive $\delta s$ leave the next minimum within reach. Increment to $s=1$ directly.
    \item $0\leq\delta s\leq\delta s_\mathrm{min}$ for a set $\delta s_\mathrm{min}$: progress via warm starts is impractically slow. Increment by $\delta s_\mathrm{min}$.
    \item $\delta s>\delta s_\mathrm{min}$: increment by $\delta s$.
\end{itemize}
In our applications, $\delta s_\mathrm{min}$ will be set equal to the increment suggested by Eq.~\eqref{eq:s_from_v}, but other choices may also be considered.

\subsection{Computational complexity of the algorithm}

To assess the overall cost of our algorithm and compare it to that of a conventional VQLS, it is useful to summarize it here and take a look at its parts separately. At each cycle of the adiabatic evolution, we perform the following operations:
\begin{itemize}
    \item Compute the expectation values of the operators $\hat{a},\hat{b}$ and $\hat{c}$ defined in Eq.~\eqref{eq:abc} on the quantum circuit, both for the current minimum $\theta^*$ and for the shifted parameter values necessary to compute the Hessian $\mathcal{H}[C_s(\theta^*)]$ at the same point. The complexity of this operation is $\mathcal{O}(dnM)$, where $M$ is the maximum number of operators in the Pauli decomposition of $\hat{a},\hat{b}$ and $\hat{c}$.
    \item If the Hessian has no negative eigenvalues, find $\delta s$ such that the smallest eigenvalue of $\mathcal{H}[C_{s+\delta s}(\theta^*)]$ changes sign, using a classical, Newton-like root-finding method and standard diagonalization routines. The computational cost of this operation is $\mathcal{O}(\text{Poly}(dn)\log(\log(1/\epsilon)))$, where $\epsilon$ is the tolerance on $\delta s$.
    \item Increment $s$ by $\delta s$ and find the minimum of the cost function in the new landscape, using the previous minimum $\theta^*$ as a warm start. This part of the total cost depends on the particular optimization routine chosen, but it is generally quite efficient if the previous steps successfully identified a path preserving the convexity of the parameter space.
\end{itemize}
The cycle is repeated (at most) a set number of times $T$, which is arbitrary in principle but should be chosen with care, like the circuit depth $d$ or the tolerance $\epsilon$, considering the problem at hand. These hyperparameters concur in determining the performance of the algorithm as well as its cost and there is no intuitive way to determine their optimal values for any given application up front.

The circuit depth and the form of the ansatz determine the parameter space and hence the set of accessible quantum states, establishing the maximum accuracy achievable by the algorithm for the problem given. The cost function landscape, in combination with the matrix $A$, in turn entails a maximum step size in order for the convexity to be preserved by the adiabatic step. This must be guessed by setting an appropriate number of steps $T$: while the optimal $T$ is not obvious, test runs of the algorithm itself can provide some hints. On one extreme, a very low $T$ will cause the cost function to change too much for the previous minimum to be of any use in the search for the new one, resulting in effectively independent searches with no real warm starts. Vice versa, very short steps suggested by setting a high value for $T$ will most likely be overridden by the calculation of the Hessian, leading to a much lower effective cost than $T$ times the cost per step. This means that the warm-start approach is successful in optimizing the adiabatic increments, and also that setting $T$ to values higher than a certain problem-dependent limit will have no impact on the actual cost.

In the next section, we will demonstrate this algorithm on a discretized one-dimensional heat flow equation and analyze its performance and results for this particular test case.

\section{\label{sec:III} The heat flow equation}

To address the problem of solving a linear, one-dimensional differential equation with a variational quantum solver, we consider the stationary state of the in-homogeneous non-uniform heat flow equation
\begin{equation}\label{eq:heat_flow}
   \frac{\partial}{\partial t}f(z,t) = \frac{\partial}{\partial z}\bigg{(}\lambda \frac{\partial f(z,t)}{\partial z}\bigg{)}+Q_r(z),
\end{equation}
where $f$ is the temperature distribution, $Q_r$ is the heat production rate and $\lambda$ is the thermal conductivity of the medium. Usually, $\lambda$ is a function of the position $z$ and of the temperature $f$ itself, and thus the general equation is nonlinear. However, in many cases of physical interest, including in applications to geological modeling~\cite{basin_modeling,ma_case_2025,pasquale_heat_2014}, one may assume a temperature-independent thermal conductivity.

\begin{figure*}
    \centering
     \subfloat[]{\includegraphics[width=0.52\linewidth]{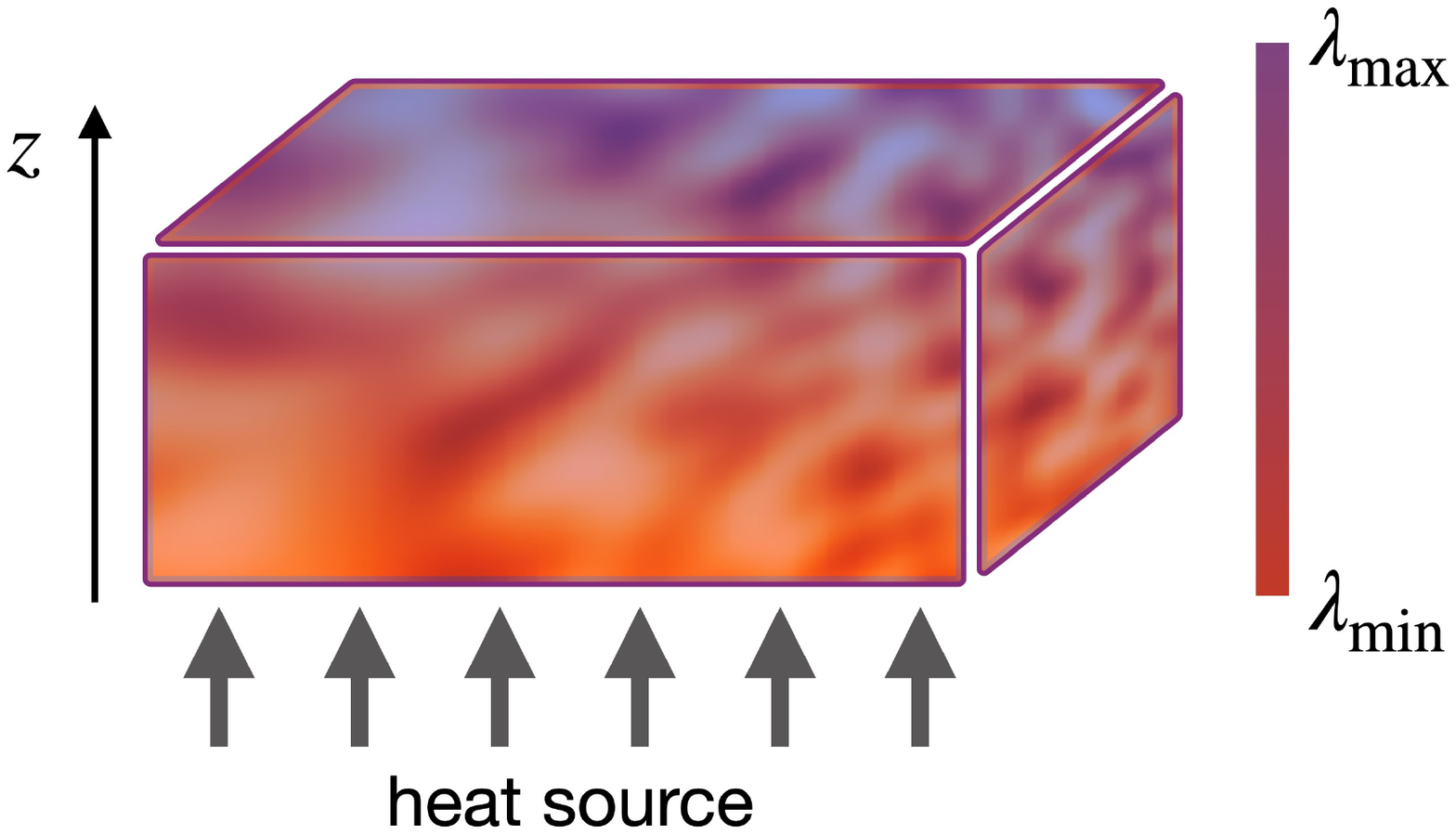}}
     \subfloat[]{\includegraphics[width=0.46\linewidth]{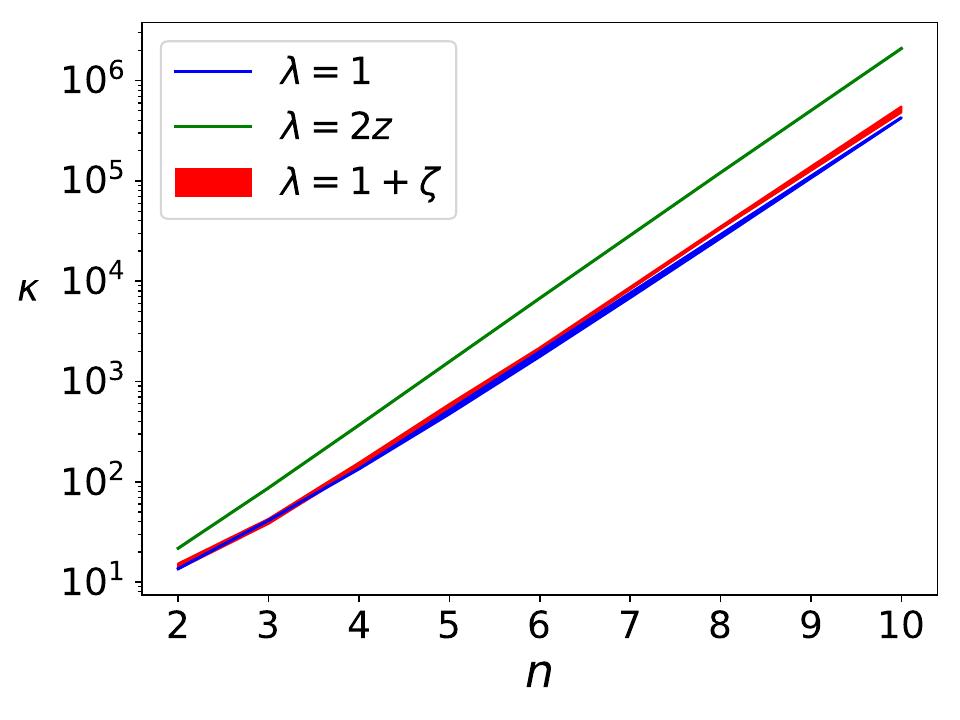}}
    \caption{(a) Representation of a block of some heterogeneous material with conductivity increasing along the $z$ coordinate up to some noise $\zeta$: $\lambda(z) = c z+\zeta$. A heat source located at the bottom of the bulk, where $\lambda(z)=\lambda_\mathrm{min}$, is shown and can be represented in Eq.~\eqref{eq:stationary} by a pointlike source term $Q_r(z)\propto\delta(z)$. Our goal is to find the stationary state of the heat flow equation for this problem on a finite lattice. (b) Condition number $\kappa$ of the matrix $A$ for the heat flow equation in the discretized form given in Eq.~\eqref{eq:heat_matrix}, for the cases $\lambda(z)=1$ (blue line), $\lambda(z)=1+\zeta(z)$ with $\zeta$ normally distributed with $\sigma = 0.2$ (red line) and $\lambda(z)=2z$ (green line). $\kappa$ increases exponentially with the number of qubits for all the analyzed cases.}
    \label{fig:4}
\end{figure*}

The stationary state is obtained by setting the left-hand side of Eq.~\eqref{eq:heat_flow} to zero and solving 
\begin{equation}\label{eq:stationary}
    \frac{\partial}{\partial z}\bigg{(}\lambda(z) \frac{\partial f(z)}{\partial z}\bigg{)} = -Q_r(z).
\end{equation}
In order to solve the problem for a system of length $L$, we first discretize Eq.~\eqref{eq:stationary} into an $N$-site regular lattice with spacing $\Delta z = L/N$, thus transforming it into a linear problem of the general form Eq.~\eqref{eq:classical_linear_system}. The resulting matrix $A$, for open boundary conditions, is the $N\times N$ tridiagonal matrix with components

\begin{eqnarray}\label{eq:heat_matrix}
   A_{i,i} & = &  -\frac{2\lambda_i}{\Delta z^2}\\
   A_{i,i-1} & = & \frac{1}{\Delta z^2}\bigg{(}\lambda_i+\frac{\lambda_{i+1}-\lambda_{i-1}}{4}\bigg{)}\nonumber\\
    A_{i,i+1} & = & \frac{1}{\Delta z^2}\bigg{(}\lambda_i-\frac{\lambda_{i+1}-\lambda_{i-1}}{4}\bigg{)}\nonumber
\end{eqnarray}
for $i=1,\dots,N$, $\lambda_i=\lambda(i\Delta z)$, and the discretized source term directly yields the components of the vector $\mathbf{b}$ as $b_i=-Q_r(i\Delta z)$. Note that we are interested in discretizations with $N=2^n$, so each added qubit doubles the spatial resolution of the discretized problem. However, it should also be kept in mind that increasing $n$ has a direct impact on the condition number $\kappa$ of matrix $A$, which increases exponentially in $n$ for this type of problem, as shown in Fig.~\ref{fig:4}~(b). This scaling of the condition number in such finite-differences problems is a known obstacle for any numerical solution method~\cite{moin_fundamentals_2010} and will be shown to affect the quality of our solutions as well.

We will now proceed to test the adiabatic quantum variational algorithm on several instances of this linear problem, defined by the form of the conductivity function $\lambda(z)$ and of the source term $Q_r(z)$.

For the conductivity, we will first consider a homogeneous material, characterized by $\lambda(z) = \lambda_0$. Next, we will include the presence of impurities in our physical model by adding a noise term, so that $\lambda(z)=\lambda_0+\zeta$, where $\zeta$ is a normally distributed random variable with zero mean and standard deviation $\sigma$. Finally, we will consider a different system in which the conductivity increases linearly with the position: $\lambda(z) = \frac{2z}{L}\lambda_0$, with or without added noise. Such a linear increase in conductivity is a common condition in geophysical settings, in which a uniform conductivity variation with depth and pressure is typically observed~\cite{ma_case_2025,pasquale_heat_2014}.

As for the heat source, we will compare different cases, all decaying exponentially with respect to $z$, so that $Q_r(z)=Q_0e^{-lz/L}$ with $l\geq 0$. In particular, we will let $l$ take on different values and compare the effect of a steeper or slower decay, including the extreme cases $l=0$, i.e.\ a constant source term $Q_r(z)=Q_0$, and $l\rightarrow\infty$, i.e.\ a point source $Q_r(z)\propto\delta(z)$. In practice, we will consider either $\mathbf{b}=Q_0\mathbf{e}_1$ for the pointlike case, or
\begin{equation}\label{eq:heat_source}
    b_{lj}=Q_0e^{-jl\Delta z/L}
\end{equation}
with $l$ chosen between $0,2$ and $5$. As discussed at the end of Section~\ref{ssec:AVQLS}, in the latter cases we will always perform a change of basis via the Householder transformation Eq.~\eqref{eq:householder_matrix} before feeding the problem into our variational algorithm, so that the quantum circuit is always initialized in the state $\ket{0}=\ket{b}$. A pictorial representation of a bulk of non-homogeneous material, with noisy conductivity along the vertical direction and a pointlike heat source, is shown in Fig.~\ref{fig:4}(a).

For the purpose of numerical solution, we choose units such that $L=1$, $\lambda_0=1$ and $Q_0=1$. All results presented in the following subsections are dimensionless.

\subsection{General results}

\begin{figure*}
    \centering
    \subfloat[]{\includegraphics[width=0.32\linewidth]{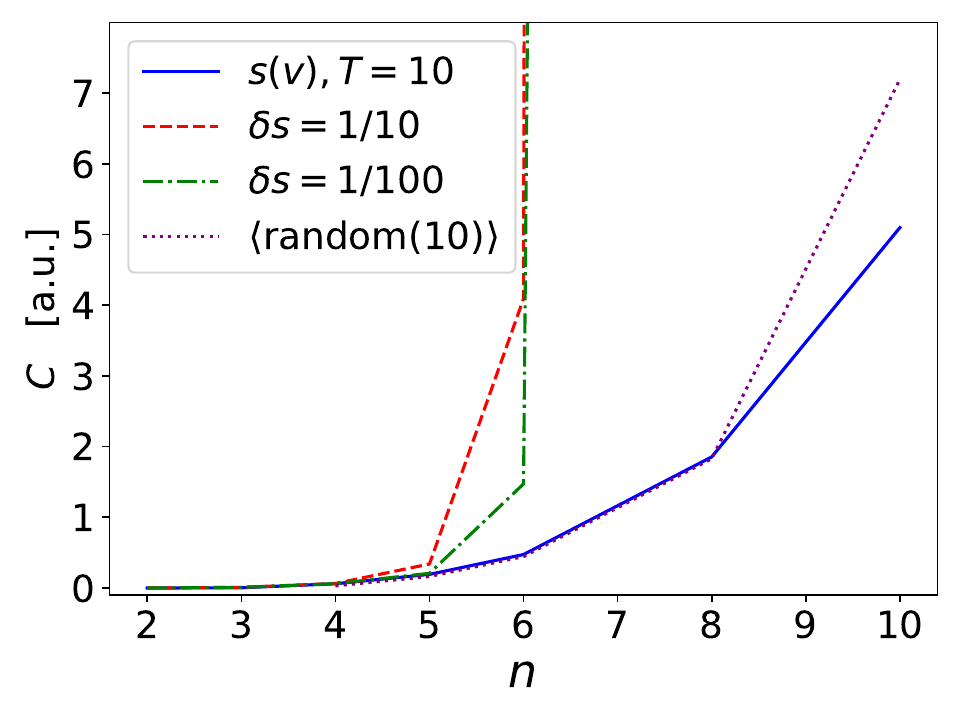}}
    \subfloat[]{\includegraphics[width=0.32\linewidth]{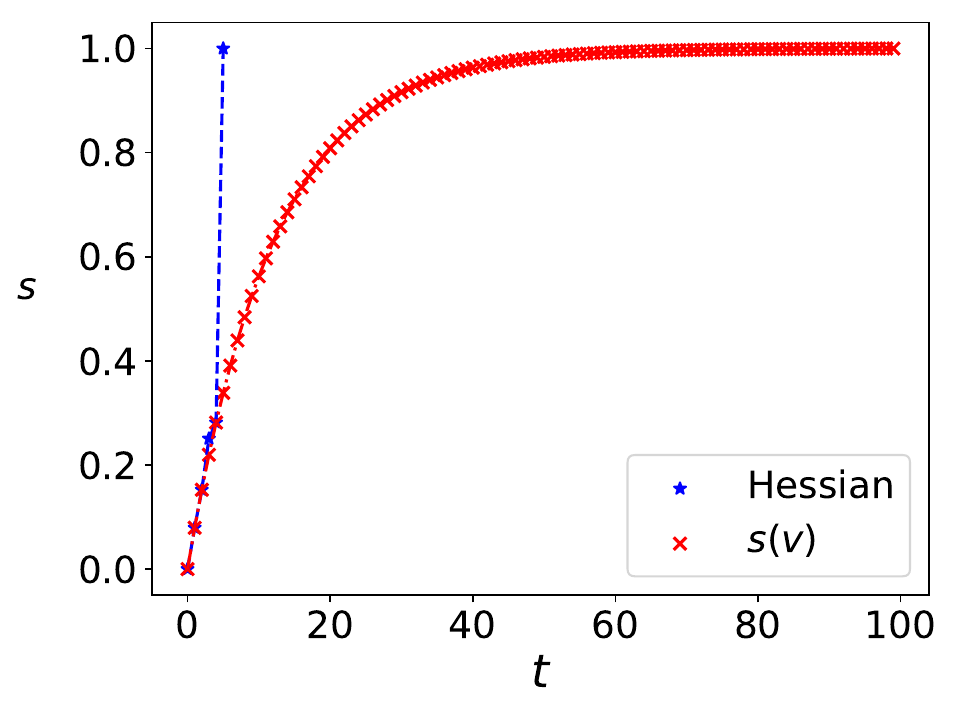}}
    \subfloat[]{\includegraphics[width=0.32\linewidth]{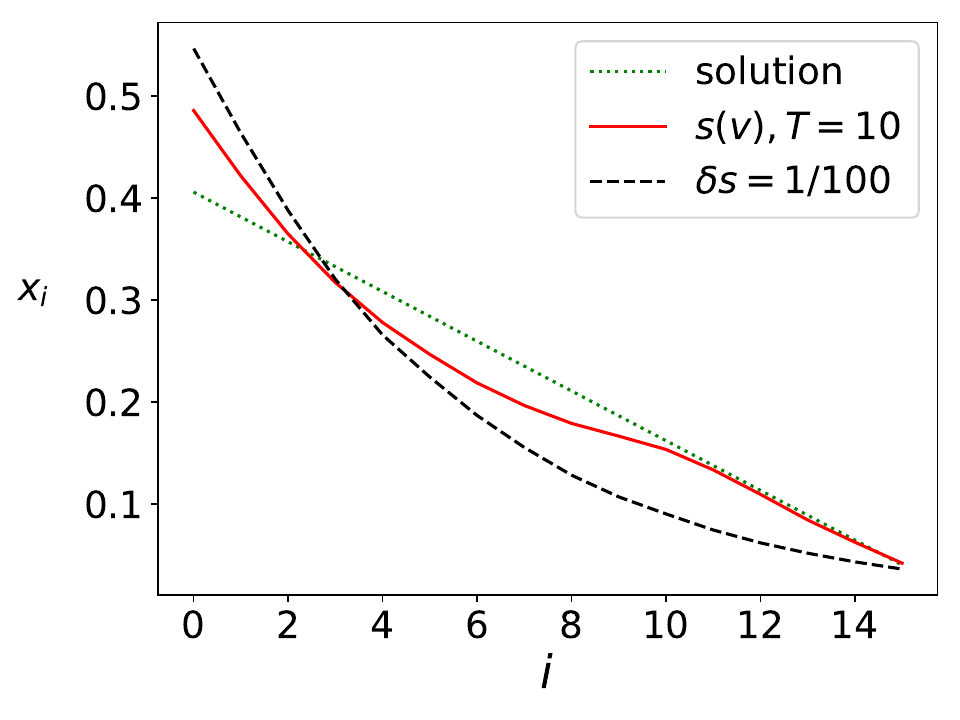}}
    \caption{A comparison of the different variational strategies considered in this paper, applied to the heat equation Eq.~\eqref{eq:stationary} with $\lambda=1$, $\mathbf{b}=\mathbf{e}_1$ and an ansatz of depth $d=1$. (a) Minimum value of the cost function for all qubit numbers $n$ considered, obtained using the adiabatic approach $s(v)$ with 10 timesteps (solid blue line), fixed $\delta s$ equal to 0.1 (dashed red line) and 0.01 (dot-dashed green line) and the best direct optimization at $s=1$ obtained from 10 random initialization points, averaged over 100 trials (dotted purple line). (b) Number of adiabatic steps needed to reach $s=1$ using the convexity analysis (blue stars) or $s(v)$ with $T=100$ (red crosses). The resulting minimum is the same. (c) Components $x_i$ of exact and variational solutions using constant $\delta s=0.01$ (dashed black line), $s=s(v)$ with $T=10$ (solid red line), and exact solution (dotted green line).}
    \label{fig:general_results}
\end{figure*}

We applied the algorithm to the problem for various configurations of the quantum circuit, differing in number of qubits $n$ and depth of the ansatz $d$, emulated using the Qiskit library. We used optimization routines provided by the Scipy package and chose the limited-memory, bounded Broyden--Fletcher--Goldfarb--Shanno (L-BFGS-B) method to search for the minima. Since the maximum system dimension considered is $N=2^{10}=1024$, which can easily be solved numerically by standard linear algebra libraries, we also used the numerical solution $\ket{x}$ obtained this way as a quality check for our results. Knowing the numerical solution, we can compute the infidelity of the variational solution $\ket{x(1)}$,
defined as
\begin{equation}\label{eq:infidelity}
    \mathcal{I}\coloneqq 1-\mathcal{F}\big(x(1),x\big),
\end{equation}
where $\mathcal{F}\big(x,y\big)\coloneqq |\braket{x}{y}|^2 $ is the fidelity between two states $|x\rangle$, $|y\rangle$.

Note that the algorithm itself has no information on $\ket{x}$ and works by minimizing the energy as expressed by the Hamiltonian $H(s)$ instead. This amounts to maximizing $|\braket{b}{A(s)|x(s)}|$ and hence the accuracy of the solution at $s=1$, defined as
\begin{equation}\label{eq:accuracy}
    \mathcal{A}\coloneqq\frac{|\bra{b}A\ket{x(1)}|^2}{||A\ket{x(1)}||^2}.
\end{equation}
In general, the two figures of merit may exhibit varying degrees of correlation depending on the spectrum of $H(1)$.

\begin{figure}
    \centering
    \includegraphics[width=1\linewidth]{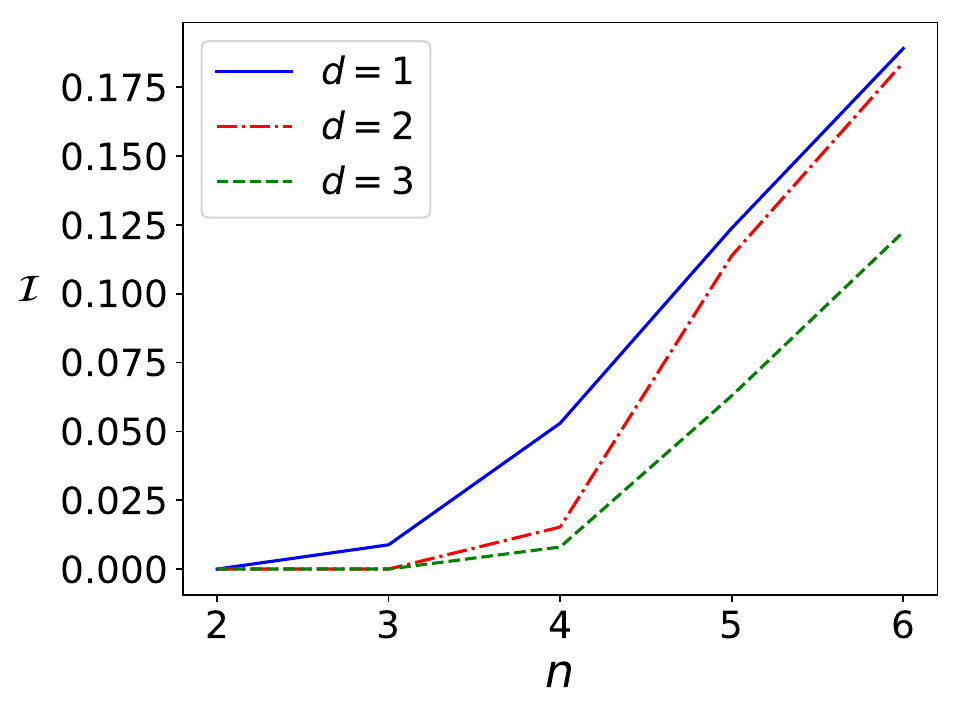}
    \caption{Infidelity $\mathcal{I}$ of the variational solution for $\lambda=1$, $\mathbf{b}=\mathbf{e}_1$, computed with $s=s(v)$ for different circuit depths. When the infidelity is around zero ($n=2$ and $3$), the depth is large enough ($d=1$ and $2$ respectively) for the image of the variational circuit to cover the whole Hilbert space and the algorithm manages to reach the exact solution.} 
    \label{fig:infidelity_with_depth}
\end{figure}

Before looking into the details of the simulations, Fig.~\ref{fig:general_results} gives a flavor of some trends emerging from several trial runs of the method for the case $\mathbf{b}=\mathbf{e}_1$, $\lambda=1$ using a circuit of depth $d=1$ and up to $n=10$ qubits, which deserve some general comments. Fig.~\ref{fig:general_results}~(a) shows the effect of different parametrization choices on the quality of the solution, as expressed by the minimum value $C$ of the cost function $C_s(\theta)$ plotted against the qubit number $n$. In particular, we show the results for $T=10$ and $100$ evenly spaced adiabatic steps, $T=10$ dynamic steps as given by Eq.~\eqref{eq:s_from_v} and sets of 10 randomly initialized direct minimizations of $H(1)$. The benefits of a problem-dependent step size can be clearly seen at $n>4$, where the quality of the solutions for $T=10$ and $T=100$ rapidly deteriorates even in this simple example with minimal circuit depth $d=1$. The random initializations (sampled 100 times in order to get a reasonable estimate of the expected quality of a set of 10 optimizations) provide a control, which turns out to be comparable to the results of the adiabatic approach up to $n=8$ qubits, but fails to keep up at higher $n$: this is a fundamental sanity check and tells us that approaching the solution adiabatically in $T$ steps is at least as efficient and accurate as performing $T$ randomly initialized searches.

In Fig.~\ref{fig:general_results}~(b) we show the prototypical effect that the analysis of the Hessian brings to the number of adiabatic steps needed to reach the final solution. When the convexity analysis suggests a negative $\delta s$, the algorithm skips all remaining intermediate steps because the current parameter values can be used as a warm start to solve the problem for $s=1$ directly. The results confirm that the quality of the solution is unaffected, since the final state obtained by performing all the $T$ adiabatic steps or by skipping to $s=1$ thanks to the Hessian check is indeed the same.

The effect of the dynamic adiabatic step is also evident if we take a look at the solution found and compare it with the numerical one: Fig.~\ref{fig:general_results}~(c) shows the vector components $x_i$ of $\ket{x}$ and $\ket{x(1)}$ computed by two runs without the Hessian check, one with a fixed $\delta s=0.01$ ($T=100$) and the other using Eq.~\eqref{eq:s_from_v} with $T=10$, for $n=4$ and circuit depth $d=2$. Again, the dynamic solution is noticeably closer to the exact one, although its overall quality is not excellent. For this result, the infidelity is $\mathcal{I}\approx 0.01$, but the accuracy is quite high ($\mathcal{A}\approx 0.9996$), signaling that the difference between $\ket{x}$ and $\ket{x(1)}$ does not translate to an appreciable difference after applying $A$. The same behavior is observed in the $\delta s = 0.01$ case, where the infidelity reaches $\mathcal{I}\approx0.07,$ with an accuracy $\mathcal{A}\approx0.9982$. This is an example of how the condition number of $A$ ($\kappa\approx 135$ in this case) may decouple the two quality measures of a solution.

Finally, it must be noted that the quality of the solutions found was consistently unsatisfactory at higher $n$, even when using the dynamic step. This is explained by the number of variational parameters $n_p=n(d+1)$ encoding the solution in the quantum circuit: in general, the state of $n$ qubits with only real gates acting on them is described by $N_p=2^n-1$ real numbers, far exceeding the expressibility of the ansatz. Fig.~\ref{fig:infidelity_with_depth} shows how increasing $d$ improves the quality of the solutions found for $n\leq 6$.

In light of the above results, we now move on to examining the performance of the full algorithm as described at the end of Section~\ref{ssec:WS}. From this point onward, the algorithm will always include the Hessian-based evaluation of the optimal $\delta s$. 

With these general observations in mind, let us now move on to the physical models.

\subsection{Case I: Constant conductivity}

\begin{figure*}
    \centering
    \subfloat[]{\includegraphics[width=0.48\linewidth]{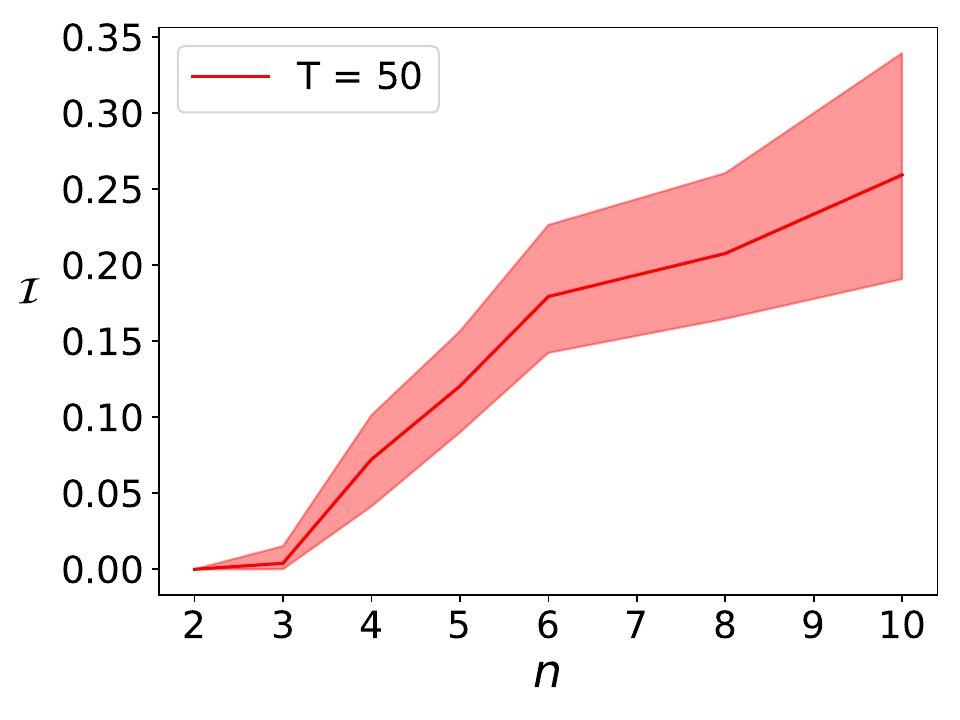}}
    \subfloat[]{\includegraphics[width=0.48\linewidth]{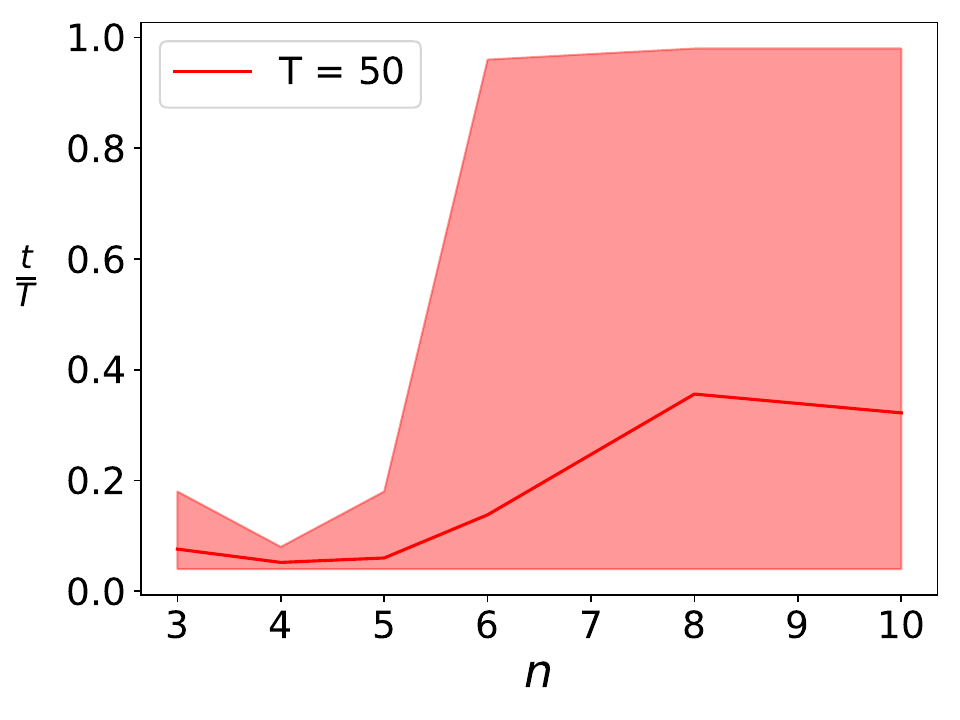}}
    \caption{Results for the $\lambda = 1+\zeta$ scenario with $\mathbf{b}=\mathbf{e}_1$, where $10$ realizations of the noise $\zeta$ with variance $\sigma=0.2$ were sampled. (a) Mean infidelity and the range between minimum and maximum infidelity (shaded area) for varying $n$. (b) Mean number of steps and the range between minimum and maximum number of steps (shaded area) for $T=50$ and varying $n$.}
    \label{fig:infidelity_noise}
\end{figure*}

In the case of a homogeneous material, we have a constant conductivity $\lambda(z) = \lambda_0 = 1$, so that $\lambda_i=1$ for all $i$ in Eq.~\eqref{eq:heat_matrix}. The lattice spacing is $\Delta z=2^{-n}$, giving matrix elements
\[
    A_{ij}=-2^{1-2n}\delta_{ij}+2^{-2n}(\delta_{i\,j+1}+\delta_{i+1\,j}).
\]
This is also the explicit form of the matrix used for the calculations shown previously.

The eigenvalues of any matrix of this particular form can be computed analytically and are given by $\alpha_k = 4\sin^2\left(\frac{\pi k}{2(N+1)}\right)$, with $k=1,\dots,N$. This implies that the condition number of $A$ grows quadratically with the number of lattice sites $N$:
\[
    \kappa=\frac{\alpha_N}{\alpha_1}=\frac{\sin^2\left(\frac{\pi N}{2(N+1)}\right)}{\sin^2\left(\frac{\pi}{2(N+1)}\right)}\sim_\infty N^2,
\]
and hence exponentially with the number of qubits $n$, as shown in Fig.~\ref{fig:4}~(b) at the beginning of this section. However, in the spirit of a more physically informed test, we added impurities to the homogeneous medium by letting $\lambda_i=\lambda_0+\zeta_i$, where $\zeta_i$ is a normally distributed random variable with standard deviation $\sigma=0.2$. The matrix $A$ is again computed from $\lambda_i$ using Eq.~\eqref{eq:heat_matrix}.

We used the adiabatic method with Hessian analysis and a maximum number of steps $T=50$ in all our calculations, after a number of tests with $T=10,100$ showed no significant differences, and performed each run with $10$ different seeds in order to average the noise. We will again consider qubit numbers up to $n=10$, but the circuit depth will be kept fixed at $d=2$ unless otherwise specified.

First, we solved the problem with a point source, i.e.\ $\mathbf{b}=\mathbf{e}_1$. The results are shown in Fig.~\ref{fig:infidelity_noise}. Fig.~\ref{fig:infidelity_noise}~(a) shows the effect of the noise on the quality of the solutions found: the results spread fairly evenly around a mean value close to the result obtained in the noiseless case (see the curve for $d=2$ in Fig.~\ref{fig:infidelity_with_depth}). The effect of the convexity check of our algorithm is shown in Fig.~\ref{fig:infidelity_noise}~(b), where we plot the number of effective adiabatic steps $t$ performed as a fraction of the full $T=50$ sequence defined in Eq.~\eqref{eq:s_from_v}: on average, we observed a reduction in the number of steps down to $t\approx 0.4\,T$, or $t\approx20$. Note that the maximum $\delta s$ compatible with the Hessian at any given $s$ is independent of $T$, so the fraction $t/T$ has no general meaning. The maximum number of steps, on the other hand, always tends to saturate the required number $T$ as the number of qubit grows: this indicates that the algorithm will often resort to the default step suggested by Eq.~\eqref{eq:s_from_v}, either because the point in parameter space is a saddle point and not a minimum (i.e.\ the Hessian has at least one negative eigenvalue) or because the $\delta s$ suggested by the Hessian is smaller than the default step, so our implementation ignores it in order to avoid getting stuck in a potentially very long chain of impractically short increments. The minimum number of steps, however, is always very close to zero, meaning that even at qubit numbers approaching $n=10$, for some realizations of the noise in $\lambda$, the Hessian analysis yielded a negative $\delta s$ and the algorithm could skip to $s=1$ directly.

\begin{figure}
    \centering
    \includegraphics[width=1\linewidth]{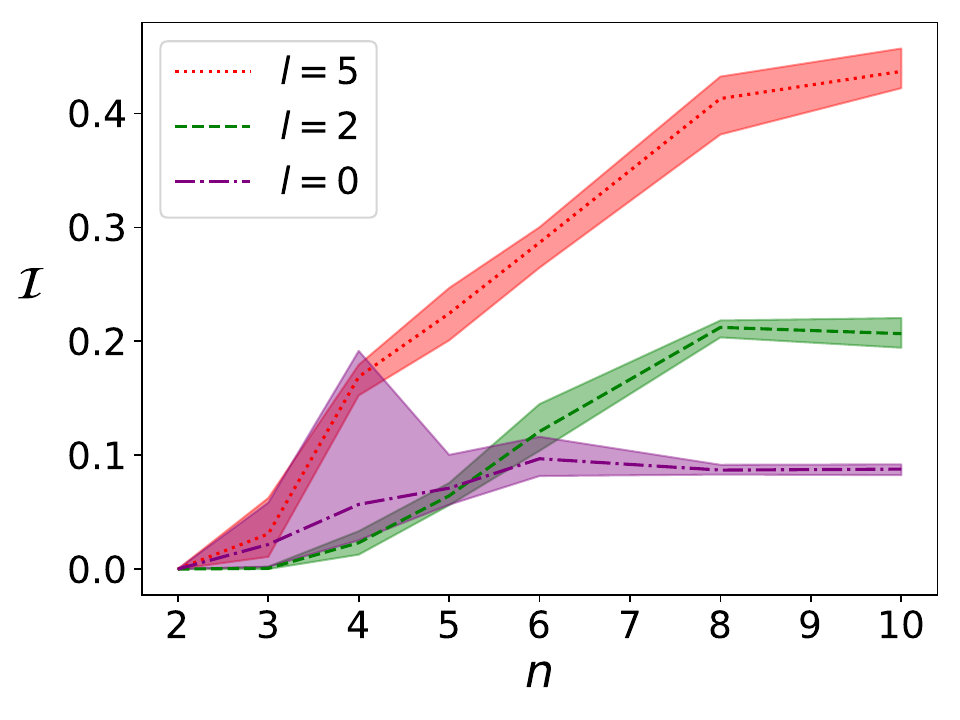}
    \caption{Mean infidelity $\mathcal{I}$ (solid line) and the range between minimum and maximum infidelity (shaded area) for different numbers of qubits $n$ in the $\lambda = 1+\zeta$ scenario with exponentially decreasing heat sources, $Q_r(z)=Q_0e^{-lz/L}$, for $l=0,2$ and $5$. The plot is the result of 10 realizations for each $n$. Note that since we plot the maximum and minimum value reached during our simulations, the shaded area is not symmetric.}
    \label{fig:infidelity_constant_noise_bl}
\end{figure}

To gain further insights into the performance of the method on this problem, we now turn to the results obtained for the family of exponential source terms $\mathbf{b}_l$ defined by their components $b_{lj}=Q_0e^{-jl \Delta z/L}$. As previously mentioned, we treat these cases by performing the Householder transformation Eq.~\eqref{eq:householder_system_equation} on the problem, so that the quantum circuit is always initialized at $\theta_i=0$, i.e.\ $\ket{b}=\ket{0}$, and all properties of the source term are encoded in the transformation applied to the matrix $A$.

\begin{figure}
    \centering
    \includegraphics[width=1\linewidth]{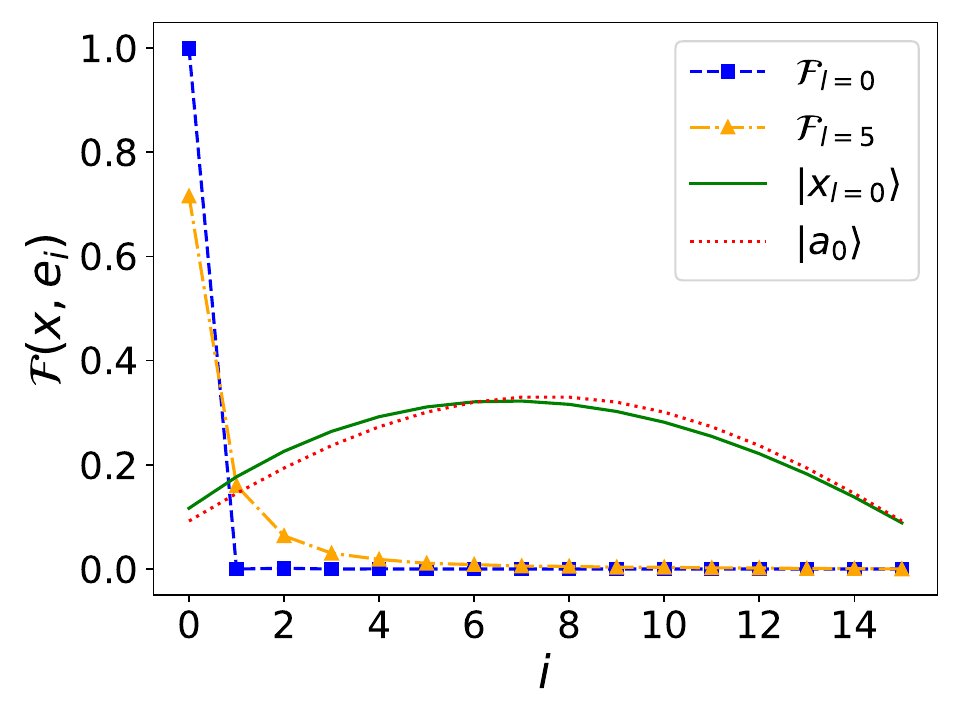}
    \caption{The fidelity $\mathcal{F}$ between the solution of the system of equations with $l=0$ (blue squares) and $l=5$ (orange triangles). The solid green line is the solution for $l=0$, while the dashed red line is the eigenvector $\ket{a_0}$ of $A$ with the lowest eigenvalue. The solutions for both $l=0$ and $l=5$ have a large overlap with the eigenvector $|a_0\rangle$, thus showing that the effective condition number is lower than the condition of the matrix $A$.}
\label{fig:overlap_states}
\end{figure}

In Fig.~\ref{fig:infidelity_constant_noise_bl}, we show the infidelity of our solutions for the cases with $l=0,2$ and $5$. Three features stand out and deserve some discussion. First, the quality of the solution consistently deteriorates as $n$ increases, confirming that the circuit depth and expressibility plays the same role as in the case of a point source. Secondly, there seems to be a correlation between better solutions and a smaller spread in infidelity, indicating that if the circuit is capable of finding a good solution its quality is not significantly impacted by the noise in the problem. Finally, we observed an interesting trend correlated to $l$: lower values of $l$ lead to better solutions, as shown by the decrease in infidelity even at higher qubit numbers. A hint towards a possible explanation for this pattern can be found by analyzing the $l=0$ case. This particular solution vector exhibits a large overlap with the eigenstate $\ket{a_0}$ of $A$ having the smallest eigenvalue, as illustrated in Fig.~\ref{fig:overlap_states}. Therefore, the solution is less susceptible to the condition number $\kappa$ of the matrix than another state with a larger overlap with higher eigenstates, since small perturbations of the solution are not amplified into large differences in the source. This allows the optimization procedure to converge quite reliably towards the desired state, thus increasing the likelihood of achieving a high-fidelity result. It also suggests that the method is not necessarily doomed to fail whenever the condition number is very high, as long as the ansatz contains the exact solution (or at least a state very close to it) and the warm-start approach correctly tracks the minimum in the parameter space. The poorer quality of the solutions with $l=2$ and $l=5$ did not come from the overlap of these states with $\ket{a_0}$, which is not significantly less than in the $l=0$ case, but have probably more to do with circuit expressibility.

\subsection{Case II: Linearly increasing conductivity}

Let us now move on to the other system of physical interest mentioned at the beginning of the section: a model with increasing conductivity $\lambda(z)=c z$, with $c>0$ a positive constant. This is a prototypical model used in geophysical science for a homogeneous vertical system, in which conductivity is proportional to pressure, and therefore to soil depth~\cite{pasquale_heat_2014}.

In this case, keeping open boundary conditions, the matrix $A$ has components
\begin{equation}
    A_{ij} = -2^{2-2n}i\delta_{ij}+2^{-2n}i\left(\frac{3}{2}\delta_{i,j-1}+\frac{1}{2}\delta_{i,j+1}\right).
\end{equation}
The condition number of $A$ exhibits the same exponential scaling in $n$ as in the constant-conductivity case already shown in Fig.~\ref{fig:4}(b). Unless otherwise specified, we will consider the medium to be not fully homogeneous by adding a noise term to the linearly increasing conductivity: $\lambda(z) = cz+\zeta$, with $c=2$ and $\zeta$ normally distributed around $0$. This time, we let $\sigma=0.05$ for the noise, since higher variances often result in the matrix $A$ not being positive and therefore physically descriptive.

\begin{figure}
    \centering
    \includegraphics[width=1\linewidth]{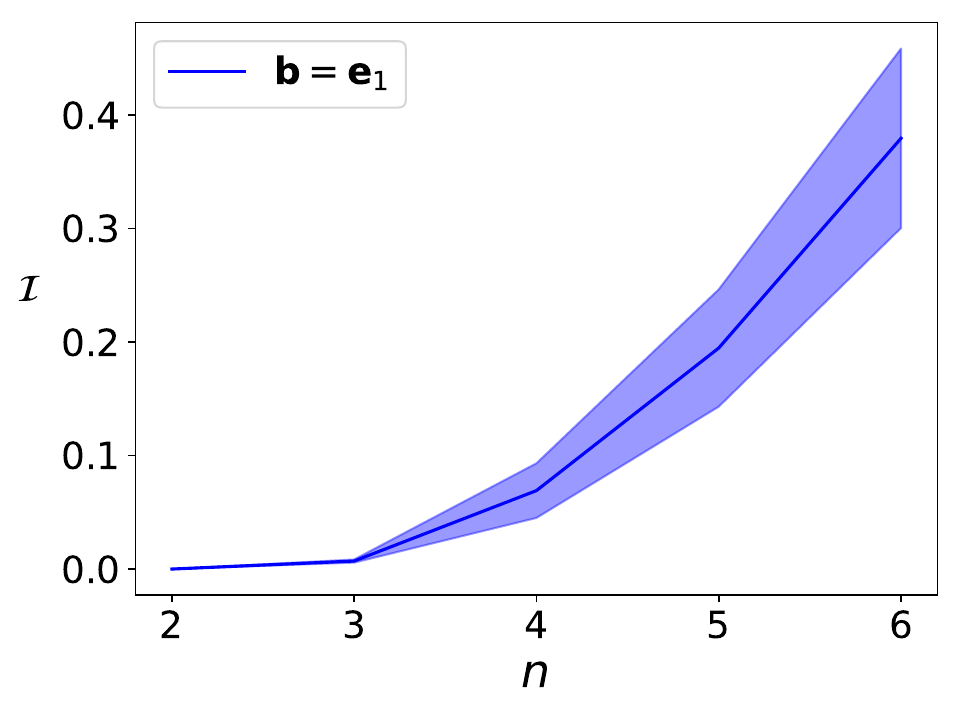}
    \caption{Mean infidelity $\mathcal{I}$ (continuous line) and the range between minimum and maximum infidelity (shaded area) for different qubit numbers $n$ in the $\lambda(z) = 2z +\zeta$ scenario with a localized heat source $\mathbf{b}=\mathbf{e}_1$. $\zeta$ is a normally distributed random variable with mean value $0$ and standard deviation $\sigma = 0.05$. The plot is the result of 10 realizations for each $n$.}
    \label{fig:infidelity_increasing_noise_var}
\end{figure}

Again, we first consider a localized heat source $\mathbf{b}=\mathbf{e}_1$ and will then move on to the exponential family defined in Eq.~\eqref{eq:heat_source}. The infidelity for the point-source scenario, plotted in Fig.~\ref{fig:infidelity_increasing_noise_var}, is generally higher than for the homogeneous case. Since the ground state in this case has a large component along $\ket{a_0}$, as we already saw in the constant-conductivity case, we can blame the larger infidelity on poor expressibility of the quantum circuit, which is evidently unable to provide the correct solution for this instance of the problem.

Switching to the exponential source terms, we find that the infidelity follows the same pattern as in the fully localized case. This happens regardless of the overlap between the solution and the first eigenstates of the matrix $A$, suggesting that the algorithm is once again struggling either to find its way towards the global minimum or to reach it given the constraints posed by the ansatz.

\begin{figure*}
    \centering
    \subfloat[]{\includegraphics[width=0.48\linewidth]{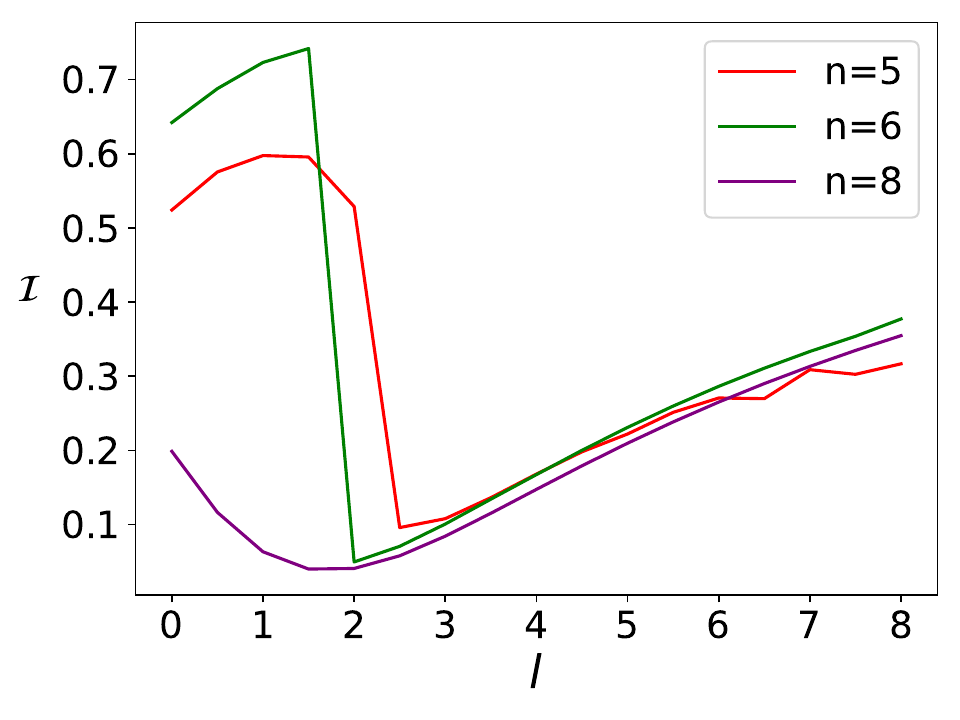}}
    \subfloat[]{\includegraphics[width=0.48\linewidth]{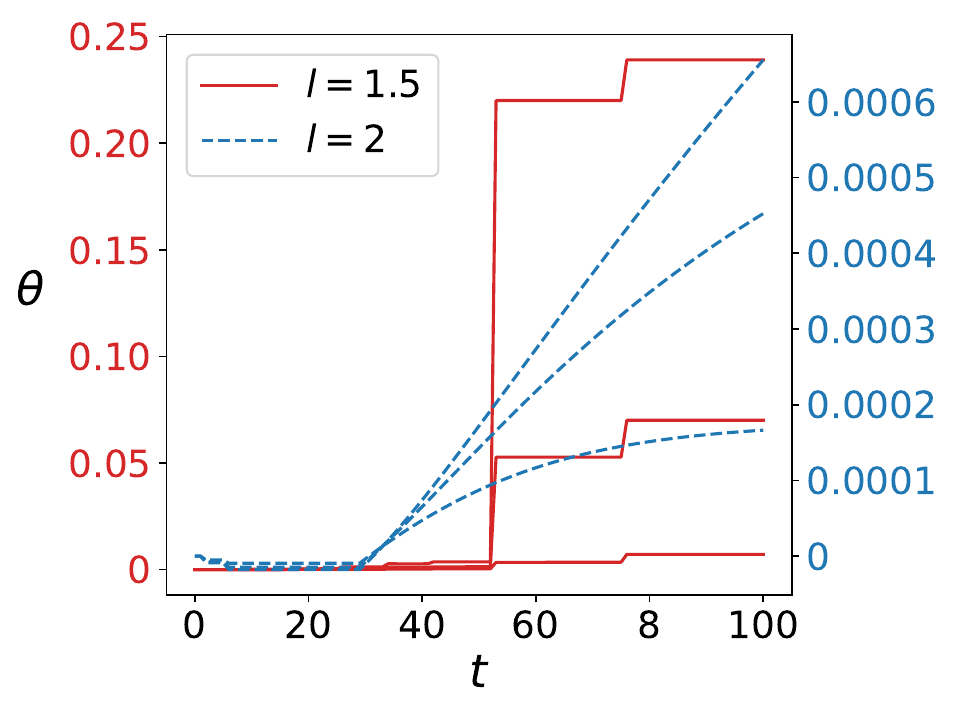}}
    \caption{Results for a heat source distribution $Q_r(z)=Q_0e^{-lz/L}$ and a maximum number of steps $T=100$. (a) Infidelity $\mathcal{I}$ as a function of the exponent $l$ of $Q_r(z)$, for $n=5,6$ and $8$. (b) Optimization paths of three selected parameters $\theta_i$ for the exponents $l=1.5$ (solid red lines) and $l=2$ (dashed blue lines) of $Q_r(z)$, for $n=6$. The angles plotted were chosen for no particular reason and the pattern is similar for all other $\theta_i$.}
    \label{fig:increasing_noise_bl}
\end{figure*}

In order to examine the link between $l$ and the quality of the solutions more thoroughly, we performed a series of runs of our algorithm letting $l$ vary between $0$ and $8$ in steps of $0.5$, removing the noise term from the conductivity and increasing the number of steps to $T=100$. In Fig.~\ref{fig:increasing_noise_bl}~(a), we show the results for $n=5,6$ and $8$. As before, we consistently found lower infidelities for lower $l$, although with qualitatively different trends for different qubit numbers. A significant transition occurs near $l=2$ for $n=5$ and $6$: the quality of the results is extremely poor for lower values of $l$ but abruptly improves above some critical value, possibly due to a different and better local minimum suddenly appearing within reach of the algorithm. Examining the parameter optimization paths for $l=1.5$ and $l=2$, we noticed a qualitative difference between the two cases, which is shown in Fig.~\ref{fig:increasing_noise_bl}~(b) for a subset of the qubit rotation angles: while for $l=2$ the adiabatic evolution appears smooth, for $l=1.5$ the parameters exhibit a piecewise evolution, with several jumps occurring as the algorithm progresses. This can be interpreted as a sign of new minima appearing outside the path traced by evolving from the initial minimum, and has the important implication that the algorithm occasionally jumped from one minimum to another, crossing the barrier separating them. Aside from the poor quality of the result despite these efforts by the algorithm, it is worth remarking that such crossings between minima are possible in small-$n$ applications such as our tests described here, but become less and less likely as the problem size increases. In fact, these barriers between independent pits and valleys in the cost function landscape can be overcome as long as the parameter space is not too large, but eventually develop into barren plateaus. In this light, it is not surprising that the result for $n=8$ seen in Fig.~\ref{fig:increasing_noise_bl}~(a) is a smooth curve: in this case, the solutions for different source terms $\mathbf{b}_l$ all lie in the same region of the parameter space even for $l\leq 2$ and the corresponding infidelity, accuracy and cost function values neatly fall in line.

\section{\label{sec:IV}Discussion and conclusion}

In this work, we proposed a modified version of the established variational quantum linear solver (VQLS) in an attempt to circumvent the notorious barren plateau (BP) problem, and used it to solve a linear differential equation in order to assess its performance. We focused our attention on the stationary solution of the heat flow equation, and applied our algorithm to several variants of the equation using different choices for the definition of the conductivity and heat source.

Our strategy to avoid the BP problem is twofold: instead of searching the whole parameter space up front, we define a parametrization interpolating between a trivial linear system and the system we want to solve, and we follow this interpolation in steps small enough for each solution to serve as a so-called warm start, i.e.\ a good guess from which to start searching for the next one. The initial convex problem, corresponding to the value $s=0$ of the adiabatic parameter guiding the transition to the target problem, always has a well known and easily prepared solution (in our implementation, the state $\ket{0}$, corresponding to all ansatz parameters $\theta_i=0$). Increasing $s$, the system and the cost function gradually change until at $s=1$ the cost function is minimized by the state $\ket{x}$ solving the linear system given. The key is to carefully determine the size of each step $\delta s$ so that the minimum of the cost function at $s+\delta s$ is easy to find starting from the one found at $s$. This is why starting from an obvious solution is essential in order for the method to work.

Our choice of $\delta s$ changes dynamically as the algorithm progresses, in order to adapt both to the system to be solved and to the behavior of the cost function at each step. A series of $T$ steps is initially created according to a rule based on the condition number $\kappa$ of the matrix $A$, but at each iteration we perform an analysis of the convexity of the cost function and use this to decide whether to move on to the next step in this series or to increase $s$ faster if it is convenient to do so.

We stress that the proposed method consistently performs better than directly optimizing the ansatz parameters for the final problem, in terms of quality of the results. This was confirmed directly by running our code with $T=1$ and finding that both our accuracy and infidelity metrics turned out worse than in the adiabatic case. However, in order to determine whether our method can indeed be considered superior to the conventional VQLS, we also need to analyze its computational complexity, since the standard VQLS involves a single optimization of the parameters, whereas our approach requires optimizing the parameters at each adiabatic step. An advantage typically emerges for systems with a large number of qubits when, by increasing the maximum number of steps polynomially with $n$ (i.e., $T\sim\mathcal{O}(n^\alpha)$ -- a problem-dependent requirement), the optimization landscape near the global minimum maintains a convex structure, which ultimately simplifies the optimization procedure required at each step. In this case, the benefits outweigh the cost associated with repeated optimizations.

As a side note, one must keep in mind that the Pauli decomposition of an arbitrary $n-$qubit operator (which is performed only at the very beginning of our algorithm, not at every step) is generally not an efficient operation. This was not discussed here because it is also a problem for conventional VQLS algorithms, but it should be noted that neither method is recommended if this decomposition cannot be carried out using a polynomial number of elementary operators. Alternatively, one might try using classical shadows techniques~\cite{boyd_training_2022,huang_predicting_2020} to approximate the result, but the quality and performance of such methods can be very problem-dependent.

\subsection*{Open points and outlook}

First of all, let us remark that in this paper we demonstrated our method only via emulator and up to a small number of qubits, $n=10$, in order to compare its results to classical numerical solutions, which are easily accessible for such problem sizes. Any performance evaluation aimed at certifying some quantum advantage would require at least twice as many qubits and real quantum hardware and was beyond the scope of this work. In addition, current NISQ devices present another series of difficulties which were not addressed here: the shot noise stemming from the limited number of measurements can overshadow the true value of the cost function, while hardware noise (such as gate errors or decoherence) can abruptly modify the landscape of the cost function, shifting the true minimum~\cite{EngDiss} and increasing the region affected by barren plateaus~\cite{Wang2021_nature}. These matters will be studied in future work, now that we have a clearer understanding of the theoretical capabilities and potential limitations of our method.

In this paper, we have explored the flexibility of our adiabatic version of the VQLS in several ways, trying to change both the form of our example problem and the hyperparameters of the method, such as the circuit depth $d$ and the number of steps $T$, in order to give a flavor of what kind of behavior to expect from it. The tests performed are by no means exhaustive, and we plan to keep exploring the potential of our algorithm both practically and theoretically, e.g. by looking for more rigorous criteria to link $T$ to the specific problem given and avoid situations in which the convexity analysis fails or makes the algorithm too inefficient.

Moreover, real-life implementations necessarily mean working with significantly larger number of parameters if the ansatz is to be sufficiently expressive for a solution to lie within its image. Like the Pauli decomposition problem, this difficulty affects any variational algorithm, but our combination of adiabatic evolution and warm starts can be very valuable in such cases, since it was proven to prevent the optimization from getting lost in barren plateaus under suitable circumstances.

It remains to be seen, however, whether this advantage comes at an acceptable cost: it is possible that, depending on the application considered, the adiabatic steps necessary to remain within a convex region in parameter space throughout the evolution are so small that the method does not outperform a standard VQLS with barren plateaus after all. We approached this question only heuristically, but it is certainly worth analyzing with more rigor, for instance by trying to link the behavior of the Hessian in the case of an adiabatically changing cost function landscape to the size of the parameter space, as done in Ref.~\cite{puig_variational_2025} for the case of unitary evolution, or to the properties of the linear problem given.
The number of parameters also drives the scaling of the convexity analysis involving the Hessian at each step in the adiabatic procedure. In order for this not to become a prohibitive cost in itself, it is necessary to choose the ansatz carefully, so that the total number of parameters is large enough for a good expressibility with respect to the problem at hand, but small enough to be easily manipulated in the calculation of the optimal step size.

Finally, we would like to remark that other areas of research, e.g. condensed matter physics or quantum chemistry, might provide much scope for testing hybrid quantum computing methods such as this variational algorithm on problems of current relevance and test the limits of such applications, including the potential advantages and limitations of present-day quantum hardware platforms.

\begin{acknowledgments}
We acknowledge financial support form the Italian National Centre for HPC, Big Data and Quantum Computing (CN00000013).
\end{acknowledgments}


\bibliographystyle{unsrt}
\bibliography{biblio}

\end{document}